\documentclass[%
 reprint,
nofootinbib,
nobibnotes,
nolongbibliography,
amsmath,amssymb,
aps,
]{revtex4-2}

\usepackage{graphicx}
\usepackage{babel,blindtext}
\usepackage{dcolumn}
\usepackage{bm}
\usepackage{xcolor}
\usepackage{soul}
\usepackage{aas_macros}

\definecolor{Brown}{RGB}{115,86,52}

\renewcommand{\lambdabar}{{\mkern0.75mu\mathchar '26\mkern -9.75mu\lambda}}

\begin{document}

\preprint{APS\textbf{}/123-QED}

\title{Dynamical Friction and Black Holes in Ultralight Dark Matter Solitons}

\author{Russell Boey}
 \email{russell.boey@auckland.ac.nz}
\author{Yourong Wang}%
\email{yourong.f.wang@auckland.ac.nz}
\author{Emily Kendall}%
\email{emily.kendall@auckland.ac.nz}
\author{Richard Easther}%
\email{r.easther@auckland.ac.nz}
\affiliation{%
Department of Physics, the University of Auckland\\ New Zealand 1010
}%

\date{\today}

\begin{abstract}
We numerically simulate the motion of a black hole as it plunges radially through an ultralight dark matter soliton. We investigate the timescale in which dynamical friction reduces the kinetic energy of the black hole to a minimum, and consider the sensitivity of this timescale to changes in the ULDM particle mass, the total soliton mass, and the mass of the black hole. We contrast our numerical results with a semi-analytic treatment of dynamical friction, and find that the latter is poorly suited to this scenario. In particular, we find that the back-reaction of the soliton to the presence of the black hole is significant, resulting in oscillations in the coefficient of dynamical friction which cannot be described in the simple semi-analytical framework. Furthermore, we observe a late-time reheating effect, in which a significant amount of kinetic energy is transferred back to the black hole after an initial damping phase. This complicates the discussion of ULDM dynamical friction on the scales relevant to the final parsec problem.
\end{abstract} 

\maketitle


\section{Introduction}

Ultralight dark matter\footnote{Also known as Fuzzy Dark Matter (FDM), Wave Dark Matter (WDM), Axion Dark Matter (ADM), or $\Psi$DM.} (ULDM) is one of many possible explanations for the non-baryonic content of galaxies \cite{PRESKILL1983127,ABBOTT1983133,Hu_2000,Lesgourgues2002ALS,Su_rez_2013,Graham_2015,MARSH20161,Ferreira_2021}. ULDM models in which the constituent scalar particle has an extremely small mass ($10^{-23}$ to $10^{-20}$ eV) exhibit wavelike phenomena on astrophysically relevant scales. This contrasts with the behaviour of traditional cold dark matter (CDM), and could provide a natural resolution to some of the noted discrepancies between CDM simulations and observations on sub-galactic scales \cite{Bullock_2017,Hui_2017,Niemeyer:2019aqm}. On cosmological scales, ULDM is indistinguishable from conventional CDM.

Very light scalar particles arise generically in a number of string theoretic approaches to particle physics \cite{Arvanitaki:2009fg}, motivating the ULDM hypothesis. In the simplest case, ULDM interactions are purely gravitational and in the non-relativistic regime the dynamics are governed by the Schr\"{o}dinger-Poisson equations. 

The characteristic ULDM halo has a distinct solitonic core embedded within a wider Navarro-Frenk-White (NFW) profile \cite{Mocz_2017}. Our focus here is on the dynamical friction experienced by a black hole moving inside the central soliton of a ULDM halo, in the absence of self-interactions. This work is motivated by interest in interactions between  supermassive black holes (SMBH) and the centres of their host galaxies. For large galaxies, the densities of the solitonic cores predicted by ULDM are very high. Consequently, dynamical friction in these regions can be expected to have a significant effect on the motion of compact objects.

Semi-analytic treatments of dynamical friction in ULDM \cite{Hui_2017} have generalised the canonical work of Chandrasekhar \cite{1943ApJ....97..255C}. Further studies have compared these models to numerical simulations  \cite{Lancaster:2019mde}. While these treatments consider dynamical friction in uniform ULDM backgrounds or those with simple velocity dispersions, the applicability of simple semi-analytic models to more general ULDM field configurations remains untested. By simulating the full dynamics of ULDM solitons when perturbed by  compact objects we aim to better understand the dynamical friction experienced by a SMBH  traversing the solitonic core of a ULDM galactic halo.\footnote{Our focus here is on ULDM models without self-interactions, but studies of dynamical friction in self-interacting models can be found in, e.g., \cite{Glennon:2023gfm, Berezhiani:2023vlo, Hartman:2020fbg, Boudon:2022dxi}.} 

This a topic of significant interest given both pulsar timing \cite{Agazie_2023} and future space-based measurements of the corresponding gravitational wave signal for  binary SMBH \cite{amaroseoane2017laser,LISA:2022kgy}. In particular, it is possible that dynamical friction from interactions with ULDM solitons may help to explain the rapid hardening of SMBH binaries to the point where gravitational wave emission can precipitate a merger. Indeed, previous studies have found that the decay of circular SMBH binaries may stall prior to the regime in which gravitational wave emission becomes dominant, resulting in predicted merger timescales which are longer than the present age of the universe. This is referred to as `the final parsec problem' \cite{Milosavljevic:2002ht}, and mechanisms through which this can be ameliorated are of great interest. Various approaches to this problem involving gas and stellar kinematics \cite{Lodato_2009,Khan_2013,Vasiliev_2015,Sesana2015} have been explored, however non-baryonic mechanisms for driving SMBH binary hardening are also a topic worthy of consideration.

To perform our simulations, we make use of {\sc PyUltraLight} \cite{Edwards_2018, Wang_2022}. This uses a pseudo-spectral Schrödinger-Poisson solver on a fixed grid to compute the ULDM dynamics, while black holes are implemented as Plummer spheres with continuously variable positions. Our test setup involves a single black hole that starts from rest at some distance from the centre of the soliton, and then falls radially inward due to gravity. The black hole then oscillates back and forth through the soliton centre, gradually losing energy due to dynamical friction. Characterising the rate at which energy can be extracted from black holes moving through ULDM solitons is of great significance, as a large amount of kinetic energy must be dissipated in order for two massive black holes to form a tight binary following a ULDM galaxy merger event. 

We compare our numerical results to simple analytic models of dynamical friction in ULDM, investigating scaling relationships between the effective stopping timescale for the infalling black hole and other system parameters. Unsurprisingly, we observe that certain key phenomena arising in the full numerical simulations are not captured by the simplified analytic models. In particular, we see that the energy injected into the system by the black hole excites the soliton, which then causes the black hole to be ``reheated'' as its motion is pumped by the time-varying ULDM field. Previous numerical investigations have revealed similar long-lasting excitations in perturbed ULDM solitons \cite{Zagorac_2022, Wang_2022}, and the present work serves to reinforce the robustness of these effects across a variety of astrophysically relevant scenarios. 

The structure of this paper is as  follows: in Section \ref{Background}  we review the theoretical description of the ULDM-black hole system and introduce {\sc PyUltraLight}, which we use to simulate this system. In Section \ref{sec:Hui_derivation}, we review a semi-analytic description of the dynamical friction experienced by a test mass moving through a uniform ULDM background, and describe how we adapt this model for our test setup. In Section \ref{Numerical Methods}, we discuss our simulation setup and the effects of various parameter choices on the evolution of the system. Section \ref{Numerical Results - Stopping Time} examines the effective stopping time for the infalling black hole, and we compare our simulations results to semi-analytic results in in Section \ref{ComparisonSection}. We conclude in Section~\ref{sec:conclusion}.

\section{Background and Overview}\label{Background}
In this work we consider the interaction of a ULDM soliton with a single infalling black hole. This system evolves in accordance with the Schrödinger-Poisson equations,
\begin{eqnarray}
    &i\hbar\dot{\psi} =-\frac{\hbar^2}{2m}\nabla^2\psi+m(\Phi_U+\Phi_{BH})\psi&,
    \label{SP1} \\[10pt]
    &\nabla^2\Phi_U =4\pi Gm|\psi|^2,&
    \label{SP2}
\end{eqnarray}
where $\psi$ is the ULDM wavefunction, $m$ the axion mass, $\Phi_U$ the potential due to ULDM self-gravity and $\Phi_{BH}$ the (Newtonian) potential due to the black hole. The black hole is accelerated due to the ULDM potential according to:
\begin{equation}
    \ddot{\mathbf{x}}_{BH}=-\nabla\Phi_U(\mathbf{x}_{BH}),
    \label{BH_evolution}
\end{equation}
where $\mathbf{x}_{BH}$ is the position of the black hole. The initial soliton profile is given by the time-independent, spherically-symmetric ground state solution of the Schrödinger-Poisson equations. This profile cannot be found analytically but is readily obtained using numerical methods. The numerical profile is itself closely approximated by \cite{Schive_2014}:
\begin{equation}   \rho(r)=\frac{1.9(m/10^{-23}\text{eV})^{-2}(r_c/\text{kpc})^{-4}}{[1+0.091(r/r_c)^2]^8}\textrm{M}_{\odot}\textrm{pc}^{-3},
\label{density_semi_analytic}
\end{equation}
where $m$ is the mass of the ULDM particle, $r$ is the radial distance from the centre, and $r_c$ is the radius at which the density drops to half its central value. 

While astrophysical ULDM halos consist of a solitonic core surrounded by an NFW-like outer region, dynamical friction is expected to be much more significant within the dense core of the soliton, so we do not include the outer halo in our analysis. However, we do note that this choice suppresses the random walk of the soliton centre of mass that is induced by interactions with the dynamical halo \cite{Schive:2019rrw,Li:2020ryg}. Other oscillations, such as ``breathing modes'' \cite{Zagorac_2022} are also suppressed at the outset of the simulation. While the interaction with the black hole can generate a variety of excitations within the soliton over time, the absence of these phenomena in the initial conditions may affect the predicted rate at which energy is extracted from the black hole. Future work will be undertaken to characterise these effects.

We use {\sc PyUltraLight} to solve the ULDM field evolution via pseudo-spectral methods. The reader is directed to \cite{Edwards_2018} for technical details. The black hole itself is implemented as a Plummer sphere, and therefore has potential
\begin{equation}
    \Phi_{BH}=-\frac{G\mathcal{M}}{\sqrt{r^2+a^2}}\, ,
    \label{BH_potential}
\end{equation}
where $a$ is the Plummer radius, $\cal M$ is the mass of the black hole, and $r$ is the distance from the centre of the black hole. The evolution of the black hole is governed by Equation \ref{BH_evolution} and takes the value of $\nabla\Phi_U$ as an input. While $\Phi_U$ is only defined at the fixed simulation gridpoints, the position of the black hole is implemented as a continuous variable. The value of $\nabla\Phi_U$ at the location of the black hole is retrieved by first calculating the gradient at each of the eight closest grid points using second order finite differencing, and then using trilinear interpolation of these values to yield the gradient at the black hole position.

We work with a fiducial ULDM mass of $10^{-21}$ eV and a fiducial soliton mass of $\text{M}_s = 5.57\times 10^7\, \text{M}_{\odot}$. Using the empirical core-halo mass relationship proposed by Schive {\em et al.}~\cite{Schive_2014}, a soliton of this size would be expected to be embedded in a halo of virial mass $6.6\times10^{10}\, \text{M}_{\odot}$.\footnote{At redshift zero the empirical core-halo relationship is $r_c = 1.6 m_{22}^{-1} \left(M_\text{vir}/10^9\text{M}_{\odot}\right)^{-1/3}\text{kpc}$. We take $m_{22} = 10$ and numerically integrate the soliton profile given in Equation (\ref{density_semi_analytic}) to obtain the total soliton mass for a given virial mass. We note, however, that this core-halo relationship is itself debated \cite{Zagorac:2022xic}.} This corresponds to a large dwarf galaxy, similar in size to the LMC \cite{2014ApJ...781..121V}. Large dwarf galaxies provide a particularly useful test bed for ULDM, as they are expected to possess wider solitonic cores than more massive galaxies like the Milky Way, and thus the novel effects of `quantum pressure' in ULDM models may be more readily observed. Furthermore, LMC-sized dwarf galaxies have also been found to exhibit optical spectroscopic signatures of central massive black holes \cite{2013ApJ...775..116R}, making this mass regime particularly relevant for the present study. Meanwhile, our fiducial black hole mass is ${\cal M}=4\times10^6M_{\odot}$. While this mass is similar to the estimated mass of the Milky Way's central black hole, it also lies at the upper end of the plausible mass range for an MBH within the LMC \cite{2017ApJ...846...14B}, such that our fiducial values of $m, \cal M,$ and $\text{M}_s$ are collectively astrophysically plausible.

\section{Semi-Analytic Model}\label{sec:Hui_derivation}

As a test mass moves through a background mass distribution, its gravitational field induces an overdense wake behind it \cite{2008gady.book.....B}. This wake then induces an effective drag force on the test mass via its gravitational field. In a ULDM fluid, this drag is automatically accounted for in a full numerical solution of the Schrödinger-Poisson system, but it is also instructive to estimate its strength semi-analytically. To do so, we consider a single point mass moving through a ULDM background with uniform density $\rho$ in the far past and a constant velocity $-v\hat{\textbf{z}}$. It is convenient to work in the rest frame of the point mass, wherein the ULDM fluid flows past with velocity $v\hat{\textbf{z}}$. If we ignore ULDM self-gravity (we will revisit the validity of this assumption in due course), this scenario is directly analogous to the scattering of a beam of particles by a Coulomb field. In this simplified scenario, the ULDM wavefunction is governed by the time-independent Schrödinger equation: $\hat{H}\psi(\textbf{r}) = E \psi(\textbf{r})$. Setting $E = \hbar^2 k^2/2m$ and $v=\hbar k/m$ we have:
\begin{equation}
    \left(\frac{mv^2}{2}+\frac{G{\cal M}m}{r}+\frac{\hbar^2}{2m}\nabla^2 \right)\psi(\textbf{r})=0,
    \label{TI_Schrodinger}
\end{equation}
where $v$ is the velocity of the ULDM flow, $\cal M$ is the point mass, $m$ is the ULDM particle mass, and $\textbf{r}$ is the position vector with respect to the point mass, with $r=|\textbf{r}|$. The analytical solution of this system is \cite{Hui_2017}:
\begin{eqnarray}    
    &\psi(\textbf{r})=\mathfrak{R}e^{ikz}\mathfrak{M}[i\beta,1,ik(r-z)]&,
    \label{wavefunction} \\[10pt] 
    &\mathfrak{R}=\sqrt{\rho}e^{\pi \beta/2}|\Gamma(1-i \beta)|,&
    \label{wavefunction_1} \\[10pt] 
    &\beta = \dfrac{G\cal M}{v^2 \lambdabar},
    \label{Beta}
\end{eqnarray}
where $\mathfrak{M}$ is a confluent hypergeometric function, $\Gamma$ is the gamma function, and $\lambdabar$ is the reduced ULDM De Broglie wavelength, $\lambdabar=\hbar/mv$. The dimensionless parameter $\beta$ represents the ratio of $\lambdabar$ to the characteristic quantum length scale associated with the point mass. The latter may be interpreted as a gravitational Bohr radius, $L_B = (\hbar/m)^2/G\mathcal{M}$, such that $\beta = \lambdabar/L_B$.

The dynamical friction force can be computed as a surface integral of the ULDM momentum flux density tensor. Taking the surface to be a sphere of radius $r$ centred on the point mass, it can be shown that the magnitude of the frictional force is \cite{Hui_2017}: 
\begin{equation}
    F_{f}=\frac{4\pi G^2 {\cal M}^2 \rho}{v^{2}} \, C(\beta, k\tilde{r}),
    \label{Hui_friction}
\end{equation}
where $\mathbf{F}= F_{f}\hat{\mathbf{z}}$ and $C(\beta, k\tilde{r})$ is the coefficient of friction given by:
\begin{equation}\label{cof}
    C=\frac{\mathfrak{R}^2}{2\beta\rho}\int\limits_0^{2k\tilde{r}}dq\big\vert\mathfrak{M}[i\beta,1,iq]\big\vert^2\left(\frac{q}{k\tilde{r}}-2-\log\frac {q}{2k\tilde{r}}\right).
\end{equation}

In Equation \ref{cof}  the value of $C$ is dependent on the value of $\tilde{r}$, and diverges for $\tilde{r}\rightarrow\infty$. Therefore, it is necessary to choose a cutoff value for $\tilde{r}$ in order to obtain a finite value of $F_{f}$. In the classical derivation of dynamical friction \cite{1943ApJ....97..255C}, this is corresponds to specifying a maximum impact parameter beyond which gravitational encounters do not contribute to the drag, and is typically taken to be of the order of the size of the system \cite{1984MNRAS.209..729T}.  For black holes orbiting within a virialised halo, $\tilde{r}$ is typically taken to be the radius of the orbit \cite{Hui_2017, Bar_Or_2019}. This cutoff makes intuitive sense, since the size of the wake generated by the travelling black hole should be of the order of the radius of the orbit, and would be expected to vanish as the orbital radius and velocity both tend to zero. However, this cutoff seems inappropriate for the scenario  here, in which the velocity of the black hole is expected to be highest near the centre. We investigate the choice of cutoff in Section \ref{ComparisonSection}.

While the integral in Equation \ref{cof} is non-trivial, it  conveniently simplifies if $\beta$ is small as we can invoke the Taylor expansion of the confluent hypergeometric function. To linear order in $\beta$ we have 
\begin{equation}
    \mathfrak{M}[i\beta,1,iq] = 1 - \beta \textrm{Si}(q) - i\beta \textrm{Cin}(q) + \mathcal{O}(\beta^2),
\end{equation}
where the usual definitions $\textrm{Si}(x)=\int_0^x\sin(t)dt/t$ and $\textrm{Cin}(x)=\int_0^x(1-\cos(t))dt/t$ apply. We may use this result to evaluate the  coefficient of friction in the limit that $\beta\ll1$:%
\begin{equation}
    C = \textrm{Cin}(2k\tilde{r})+\frac{\sin(2k\tilde{r})}{2k\tilde{r}}-1 + \mathcal{O}(\beta),
    \label{Coefficient}
\end{equation}
We discuss the regimes in which this approximation is valid in Section \ref{ComparisonSection}. 

In our adaptation of the semi-analytic approach, we replace the uniform ULDM background with the soliton profile given in Equation \ref{density_semi_analytic}. This means that the frictional force in Equation \ref{Hui_friction} becomes a function of the black hole position relative to the soliton centre. In particular, it is expected to be strongest at small radii. While this adaptation of the simple model of dynamical friction is expected to be inappropriate for high mass solitons with strongly peaked profiles, lower mass solitons with broader profiles have approximately constant density for a broad range of radii. The effects of implementing a soliton profile rather than a constant background are expected to be minimal, given  a suitable value of $\tilde{r}$.

To complete our adaptation of the semi-analytic model, we include an attractive force, $F_{a}$, directed toward the centre of the soliton. This is simply determined by the mass enclosed, $M_\text{enc}$ through:
\begin{equation}
    F_{a}=\frac{G{\cal M}M_\text{enc}}{r^2},
    \label{Newtonian_force}
\end{equation}
where $r$ is the separation between the black hole and the soliton centre. $M_\text{enc}$ is determined by numerical integration of the soliton density profile. 

We add the attractive contribution from Equation \ref{Newtonian_force} to the dynamical friction contribution from Equation \ref{Hui_friction} to compute the total force on the black hole. Numerical integration using the fourth-order Runge–Kutta method then allows us to estimate the velocity and position of the black hole over time. Some care must be taken to avoid integration errors as $r,v\rightarrow0$, so cutoffs are imposed at $r = 0.1$ pc and $v = 0.1 \text{ms}^{-1}$, respectively.

\begin{figure}[tb]
    \centering  \includegraphics[width=\linewidth]{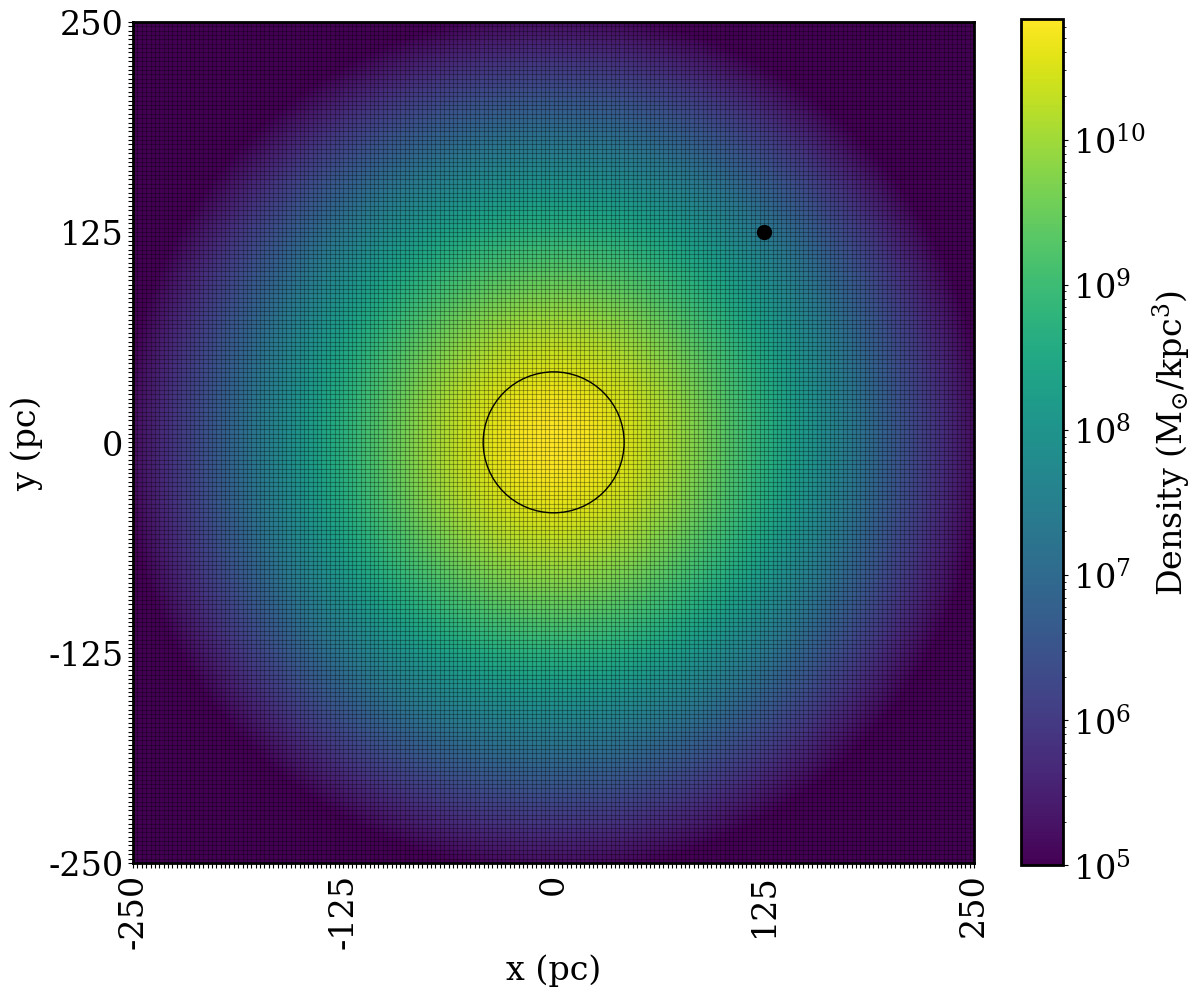}
    \caption{A zoomed-in view of the initial setup for numerical simulations. The position of the black hole is indicated by the black point, while the central circle represents the soliton's core radius $r_c$. The density distribution of the soliton is identified by the colour bar. Note that we show only the innermost $0.5$ kpc of the simulation region, though the entire box has a side length of $2$ kpc.}
    \label{setup}
\end{figure}

When invoking the soliton profile to compute both the attractive and frictional forces, we assume that it is undisturbed by the motion of the black hole. This is a reasonable assumption for black holes which are relatively small compared to the host soliton, but becomes less suitable as the black hole mass increases. As such, we expect poorer agreement between the adapted semi-analytic model and numerical simulations when the black hole mass becomes larger. Indeed, it has been previously observed that perturbed solitons tend to develop persistent `breathing modes' which will have a non-trivial effect on dynamical friction for larger perturbers \cite{Zagorac_2022,Widmark:2023dec}.  

\section{Numerical Methods}\label{Numerical Methods}

Our numerical simulations are initialised with a black hole at rest in the x-y plane, and an initially undisturbed soliton at the centre of a $2\text{ kpc}^3$ box, as shown in Figure \ref{setup}.\footnote{We repeated our fiducial simulations in a modified {\sc AxioNyx} \cite{Schwabe_2020}, and the results were sufficiently similar to confirm that we can work entirely within the simpler {\sc PyUltraLight} in this paper.} The initial coordinates of the black hole are $(0.125, 0.125, 0)$ kpc for all runs, such that the black hole lies outside the solitonic core radius at the onset of the simulation for all parameters we consider. We choose a position that lies on a diagonal to increase the distance between the black hole and the edge of the box, reducing spurious forces from periodic boundary conditions. 

We note that the black hole traverses several orders of magnitude in ULDM density as it travels through the soliton. We therefore do not expect a precise correspondence between the numerical results and the semi-analytic model, in which the initial background density is perfectly uniform. However, the ULDM density does become approximately constant within the solitonic core radius, and it is within this region that the dynamical friction is expected to be most significant. Hence, a comparison to the semi-analytic model remains well-motivated, particularly at small radii.\footnote{Intuitively, we can consider a crude model of the soliton as a ``top hat'' density profile, for which density is too low in the outer regions for dynamical friction to be significant, while within some core radius radius the density is both high and approximately constant. This approximation is most suitable for low-mass solitons.}

\begin{figure}[tb]
    \centering
    \includegraphics[width=\linewidth]{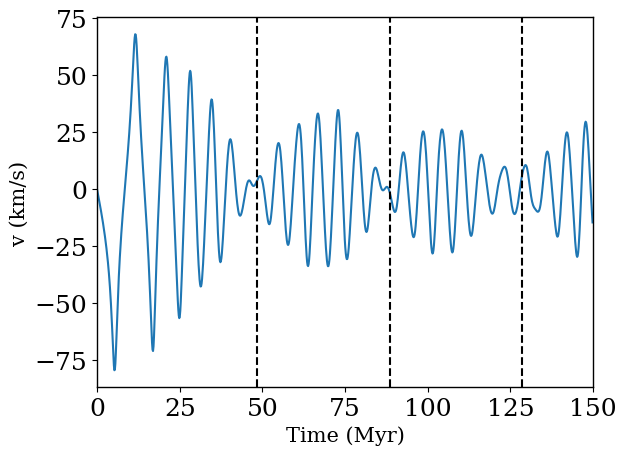}
    \caption{Evolution of the black hole velocity for our fiducial simulation. A reheating phase begins at $\sim 50$ Myr, after a period of initial decay. Further reheating events are seen at $\sim 90$ and $\sim 130$ Myr, as indicated by the dashed lines.}
    \label{Velocity_example}
\end{figure}

\begin{figure*}[tb]
    \centering
    \includegraphics[width=\textwidth]
    {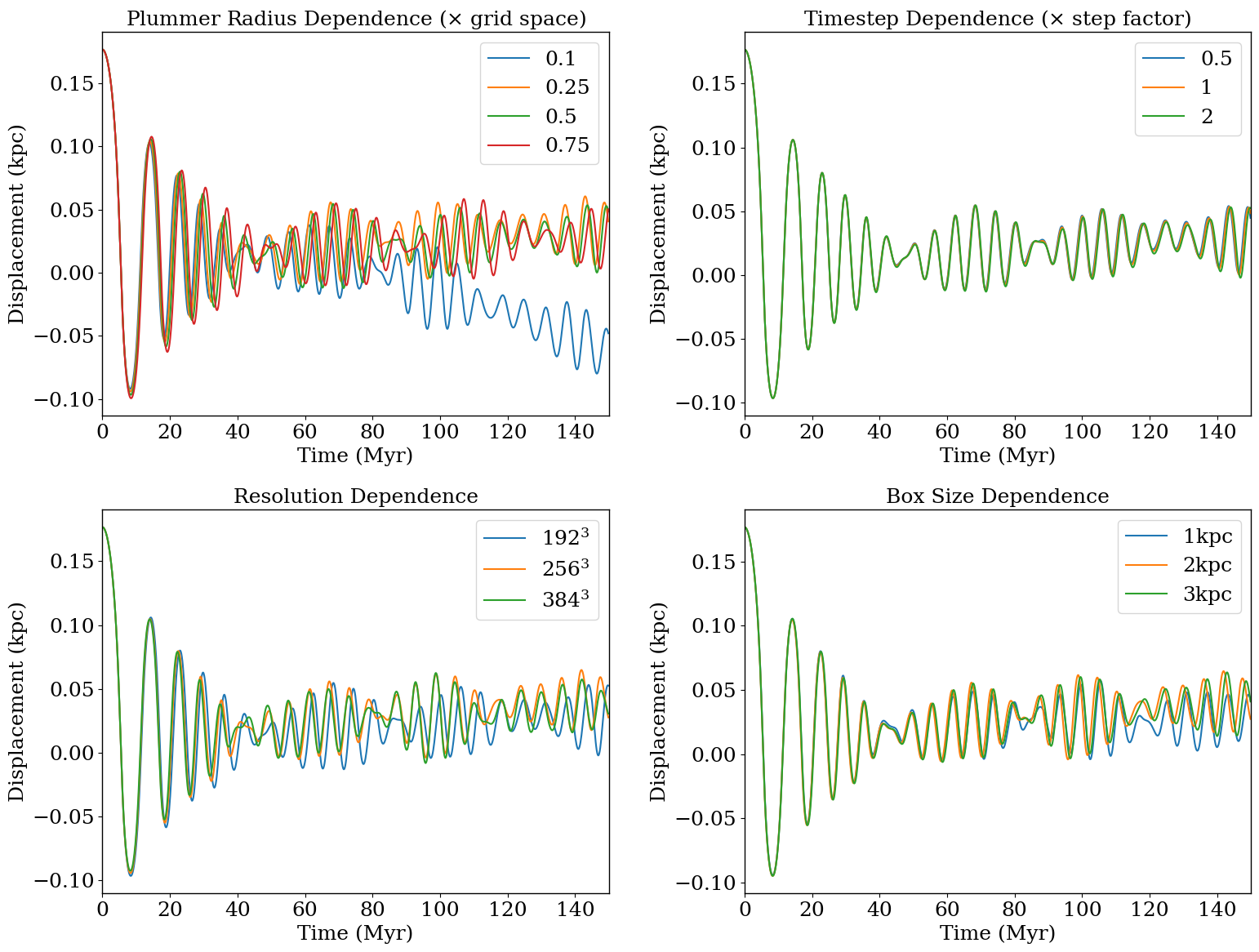}
    \caption{Sensitivity to integration parameters. Default scenario is a $192^3$ grid, 2kpc box size, and a stepfactor of 1, relative to the CFL condition \cite{Edwards_2018}. Top left: Dependence on the Plummer radius.   Top right: Dependence on the timestep. Bottom left: Dependence on resolution. (The $384^3$ run has a stepfactor of 2 for computational convenience), Bottom right: Dependence on  box size; simulations at fixed numerical resolution; e.g. $128^3$ at 1 kpc, and $384^3$ at 3 kpc.}
    \label{combined_dependencies}
\end{figure*}

A benchmark simulation at a resolution of $192^3$ using our fiducial parameters reveals that, as expected, the black hole initially falls radially toward the centre of the soliton, and then oscillates back and forth through the centre while losing kinetic energy due to dynamical friction. Interestingly, however, this loss of kinetic energy does not continue indefinitely. Rather, the black hole undergoes repeated phases of ``reheating'', during which the amplitude of the oscillations increases, as illustrated in Figure \ref{Velocity_example}.  Physically,  the passage of the black hole excites time-varying  modes in the soliton, which in turn transfer kinetic energy back to the black hole. This periodic reheating continues throughout the simulations, without a significant decrease in the characteristic amplitude.  

To assess the sensitivity of the features in Figure \ref{Velocity_example} to simulation parameters, we perform  several verification runs,  illustrated in Figure \ref{combined_dependencies}. The key  parameters are the Plummer radius of the black hole, the spatial and temporal resolutions, and the physical box size. 

\begin{figure*}[tb]
    \centering
    \includegraphics[width=\textwidth]{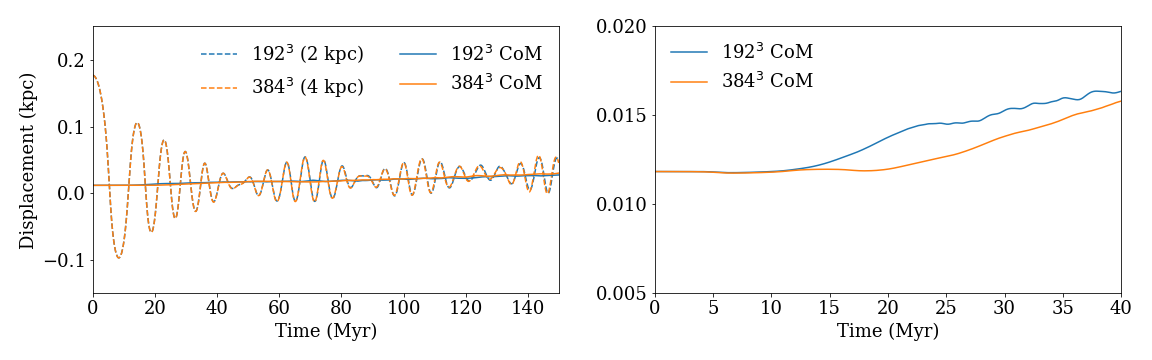}
    \caption{Plots of the centre of mass for the 192 resolution in a box size of 2kpc, as well an equivalently resolved run with 384 resolution in a 4kpc box. Both plots contain the same data, with the right plot zoomed in to more clearly demonstrate the centre of mass behaviour}
    \label{CoM_Displacement}
\end{figure*}

In the semi-analytic model, the black hole is modelled as a point mass. However, in our simulations, we implement the black hole as a Plummer sphere to avoid numerical errors that can arise when the centre of a point mass potential passes close to a grid point. Different choices of Plummer radius correspond to physically distinct scenarios, so we do not expect exact correspondence between simulations when we vary this parameter. However, we note that the key statistic, namely, the stopping time, is largely unchanged when the Plummer radius is varied between 0.25 and 0.75 grid spaces, such that a default value of $0.5$ grid spaces is suitable for the present study. By contrast, a Plummer radius of $0.1$ grid spaces proves to be too small, and numerical errors become significant in this case. This is illustrated in the top left panel of Figure \ref{combined_dependencies}. 

As discussed in Ref~\cite{Edwards_2018} the Courant-Friedrichs-Lewy (CFL) bound is a natural benchmark for the timestep in hyperbolic systems, but there is less clarity with a Schr\"{o}dinger-Poisson solver.  The {\sc PyUltraLight} `stepsize' parameter expresses the chosen timestep relative to the CFL value. The top right panel of Figure \ref{combined_dependencies} shows results for $0.5$, $1.0$ and $2.0$ -- there is no discernible difference for the two lower values and a small variation at late times with the larger value. We thus use the  {\sc PyUltraLight} default of $1.0$ for the `stepsize' parameter to balance performance and precision, unless stated otherwise.  

The sensitivity to spatial resolution is shown in the lower left panel of Figure \ref{combined_dependencies}. Clearly ``more is better'' but the computational cost rises rapidly with box size. Consequently, we default to  $192^3$ which appears to be sufficient for the initial damping phase, but larger grids (or adaptive mesh refinement \cite{Schwabe_2020}) would be needed for a detailed study of the reheating phase. 

We illustrate the impact of the physical box size in the lower right panel of Figure \ref{combined_dependencies}. Due to the periodic boundary conditions, a solition in a smaller box is affected by the gravitational potentials of the unphysical `neighbouring' solitons. Choosing a box side length from $1\text{ kpc}$ to $3\text{ kpc}$ has a minimal effect on simulation results during the initial damping phase (where physical grid resolution has been kept constant). Good convergence is observed between $2\text{ kpc}$ and $3\text{ kpc}$ even into the reheating phase but there is some deviation at later times. A side length of $2\text{ kpc}$ is chosen for our suite of simulations. 

The finite simulation volume also induces a small apparent shift in the apparent centre of mass, as shown in Figure \ref{CoM_Displacement}. Matter ejected from the soliton (especially during the first pass(es) of the black hole) can escape to arbitrary distances. However, when it crosses the boundary it ``reappears'' on the other side, biasing the centre of mass calculation. Figure \ref{CoM_Displacement} also shows that the onset of this  effect is delayed is a larger box size, confirming this interpretation.  However, the quantity of material involved is small and the central dynamics are largely unchanged. 

\section{Numerical Results}\label{Numerical Results - Stopping Time}

\begin{figure}[tb]
    \centering
    \includegraphics[width=0.45\textwidth]{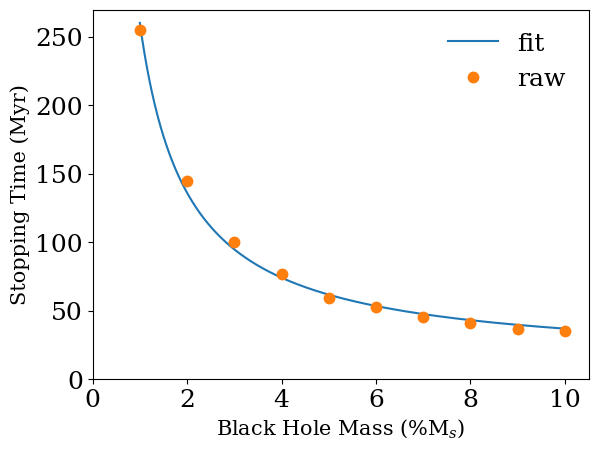}
    \caption{A fit of the stopping time to an inverse relationship with the black hole mass.}
    \label{Black_Hole_Mass_Dependence}
\end{figure}

\subsection{Black Hole Mass Dependence} 

\begin{figure*}
\centering
\includegraphics[width=0.9\textwidth]{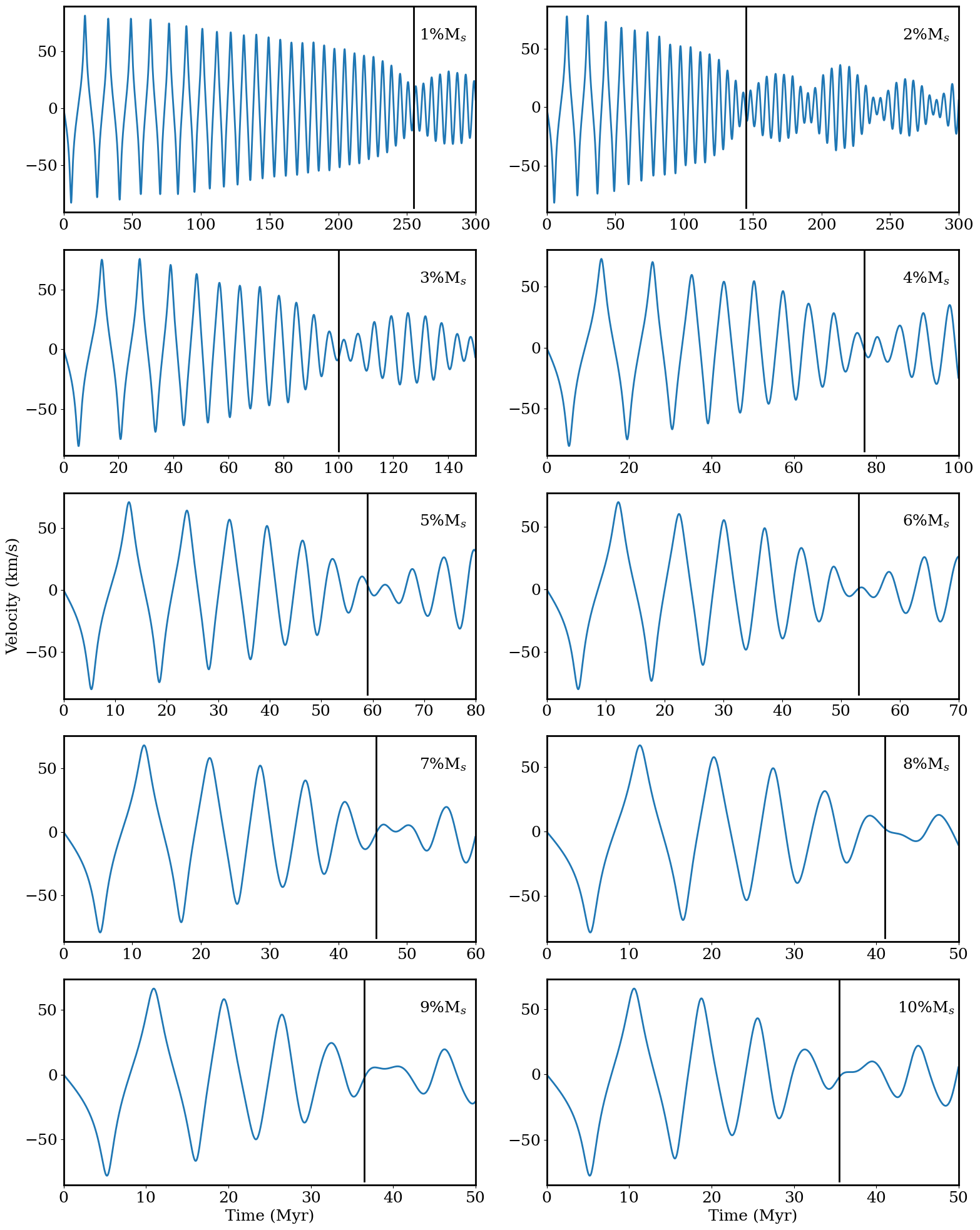}
\caption{Combined plots of the black hole velocity against time for black hole mass dependence. Black hole mass increases by $0.01 \text{M}_{s}$ from top left to bottom right, starting at $0.01 \text{M}_{s}$ and ending at $0.1 \text{M}_{s}$. The black vertical lines indicate the stopping time. Note that in this and other plots, the scales of the $x$ axes vary.}
\label{Black_hole_mass_compiled}
\end{figure*}

We first assess the dependence of stopping time on the mass of the black hole.\footnote{We define $t_\text{stop}$ as the time at which the kinetic energy of the black hole on its shortest radial pass is at its minimum. This corresponds to the point where the black hole reaches the smallest maximum radial displacement at the end of a pass. After this point, the maximum radial distance of the black hole in successive passages starts to increase as the reheating phase begins. The value of $t_\text{stop}$ is inherently imprecise, since we cannot resolve intervals substantially shorter than the period of the oscillations.} We consider masses between $0.01 \text{M}_{s}$ to $0.1 \text{M}_{s}$, with the soliton mass fixed at the fiducial value of $\text{M}_{s}=5.57\times10^7\text{M}_{\odot}$, and a ULDM particle mass of $m=10^{-21}$ eV.   Figure \ref{Black_Hole_Mass_Dependence} shows an empirical fit of the relationship 
\begin{equation}\label{fit_form}
    t_{\text{stop}}=\frac{A}{\mathcal{M}} + B,  
\end{equation}
to the raw results,  which are shown in 
Figure \ref{Black_hole_mass_compiled}. We find $A = 1.38\times 10^8\text{ M}_{\odot}\text{Myr}$ and $B=12.4\text{ Myr}$.
As expected,  the stopping time decreases with increasing black hole mass.

\begin{figure*}[tb]
\centering
\includegraphics[width=0.9\textwidth]{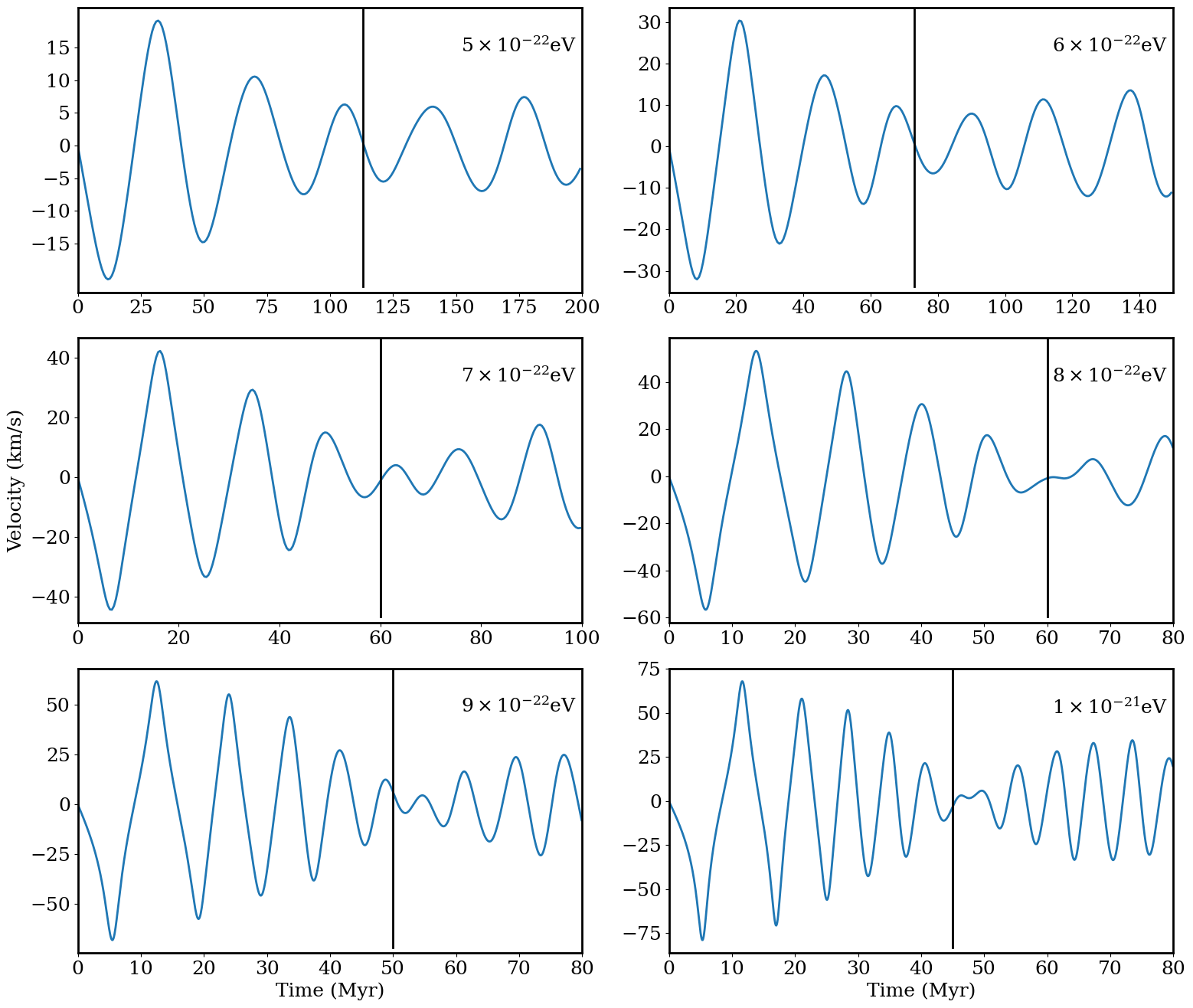}
\caption{Combined plots of the black hole velocity against time for ULDM particle mass dependence. ULDM mass increases by $10^{-22}$ eV going from top left to bottom right, beginning at $5\times10^{-22}$ eV and ending at $10^{-21}$ eV.}
\label{Axion_mass_compiled}
\end{figure*}

One can qualitatively justify the form of Equation \ref{fit_form}; each black hole is initialised with the same gravitational potential energy per unit mass. Hence,  in the absence of dynamical friction (which is  assumed to be sub-leading) the maximum velocity and infall time, $t_\text{in}$, will not depend on the black hole mass. From equation \ref{Hui_friction}  the deceleration due to dynamical friction is proportional to the black hole mass. The time taken for the dynamical friction to reduce the velocity to zero following the first infall  should  be inversely proportional to the deceleration, yielding   $(t_\text{stop} - t_\text{in})\propto\mathcal{M}^{-1}$, which can be rearranged  to give Equation \ref{fit_form}. 

\subsection{ULDM Particle Mass Dependence} 

Figure \ref{Axion_mass_compiled} shows the results of varying the ULDM particle mass, in the range $[5\times10^{-22}, 1\times10^{-21}]$ eV, in $10^{-22}$ eV increments. This does not span the ULDM masses discussed in the literature \cite{Ferreira_2021}, but is  broad enough to capture the impact on the stopping time while keeping  other simulation parameters fixed. In particular, for a given soliton mass the core radius varies\footnote{Note this is the scaling relationship for the soliton solution itself, and should not be confused with the core-halo scaling  \cite{Schive:2014hza}.} as $r_c \propto m^{-2}$, so a smaller $m$ requires large simulation volumes to fully contain the soliton. The lower end of our ULDM mass range thus arises from the need to keep the soliton away from the edges of the box, while the upper limit ensures the we resolve the de Broglie wavelength without increasing the grid resolution.

\begin{figure*}[tb]
\centering
\includegraphics[width=0.9\textwidth]{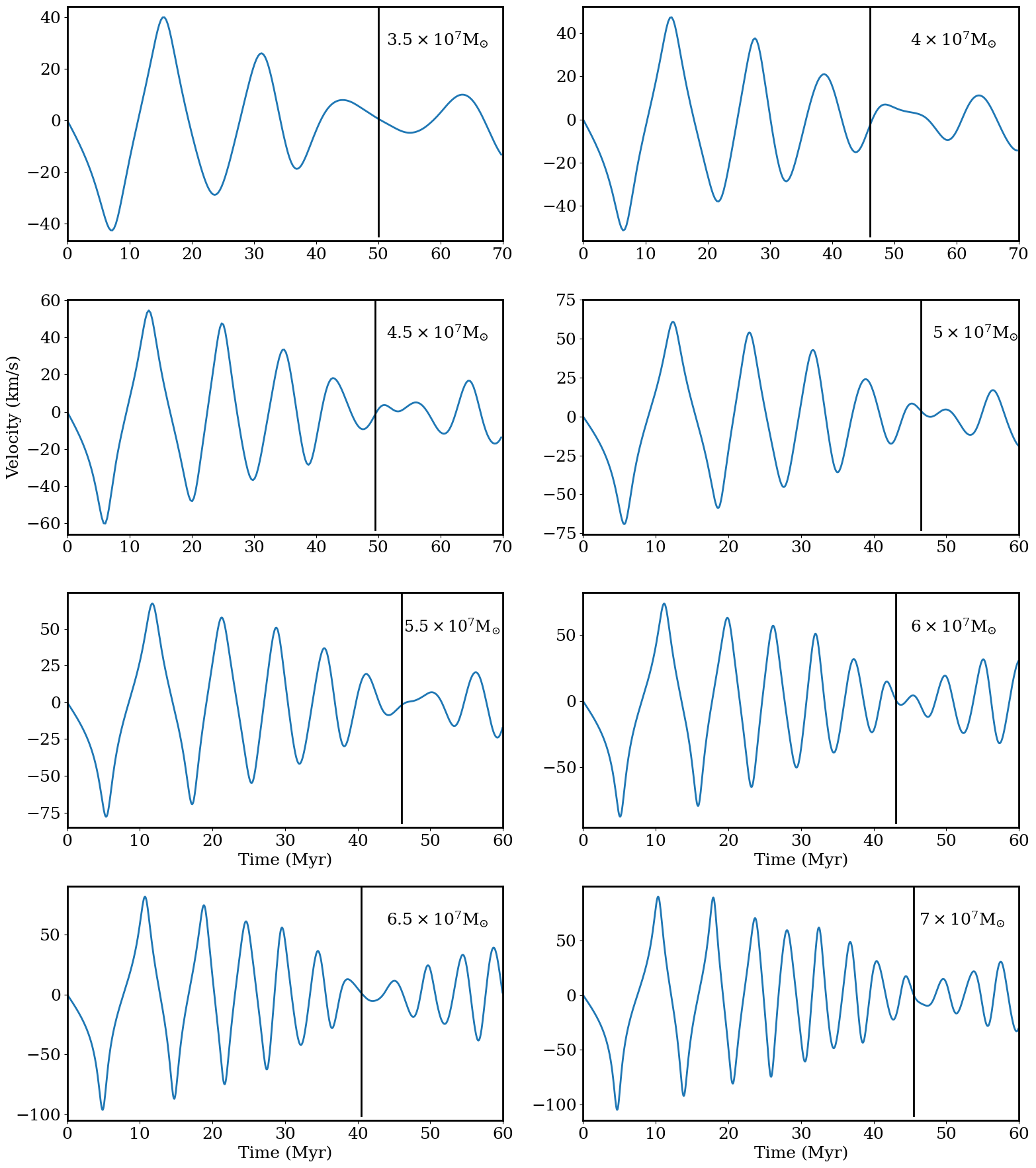}
\caption{Combined plots of the black hole velocity against time for soliton mass dependence. Soliton mass increases by $5\times10^6\text{M}_{\odot}$ from top left to bottom right, starting at $3.5\times10^7\text{M}_{\odot}$ and ending at $7\times10^7\text{M}_{\odot}$.}
\label{Soliton_mass_compiled}
\end{figure*}

The soliton is more compact at higher ULDM masses, increasing the fraction of its mass internal to the initial radial position, boosting the acceleration experienced by the black hole. The resulting increase to the central density of the soliton increases the drag; via Equation \ref{density_semi_analytic}, $\rho(r=0) \propto m^6$ and the friction is proportional to $\rho$, via Equation \ref{Hui_friction} but operates in a smaller spatial region.  Finally, changing the ULDM mass  non-trivially affects the value of $C(\beta, k\tilde{r})$.  At lower ULDM particle masses, the soliton expands, so at fixed initial separation each pass takes longer to complete and the peak velocity is reduced, although there are fewer passes before $t_\text{stop}$. The combination of these effects ensure that at lower ULDM masses, the stopping time actually increases. 

This dependence is not purely due to the change in the dynamical friction coefficient from the ULDM particle. Since the mass of the particle affects the profile of the soliton, these results are specific to the system setup that we have chosen in this work. A fixed background density would be required to test exclusively the dependence on the ULDM particle mass, although this would not be astrophysically meaningful.

\subsection{Soliton Mass Dependence}

Figure \ref{Soliton_mass_compiled} illustrates the dependence of the stopping time on  soliton masses from $3.5\times10^7 \text{M}_{\odot}$ to $7\times10^7 \text{M}_{\odot}$. The ULDM and black hole masses have their fiducial values and the lower end of the soliton mass range is chosen so that the black hole mass  does not significantly exceed $10\%$ of the soliton mass. 

There is no clear trend in $t_\text{stop}$, with all values in the range 40-50 Myr. Again there are competing effects in play. A decreased soliton mass lowers the black hole velocity, due to the decreased depth of the soliton potential. This is amplified by the core radius being inversely proportional to the soliton mass so that a greater proportion of the soliton lies outside the initial radial position of the black hole. However, the central  density scales as $\rho \propto M_{s}^4$,  reducing the dynamical friction near the centre of the lower mass solitons. Additionally with lower soliton masses each pass takes longer, but fewer are needed to reach the stopping time. 

\section{Comparison to Semi-Analytic Model}\label{ComparisonSection}

When comparing our numerical results to the semi-analytic model of dynamical friction  from  Section \ref{sec:Hui_derivation}  a number of limitations must be accounted for.\footnote{It goes without saying that the dynamical friction can not explain the ``reheating'' of the black hole, and in this Section we focus on the initial phase, prior to $t_\text{stop}$.} Firstly, the expression for the coefficient of friction in Equation \ref{Coefficient} assumes that $\beta\ll 1$. However, for our fiducial  parameters we find
\begin{equation}    
\beta\approx 9\left(\frac{\mathcal{M}}{4\times 10^6\text{M}_{\odot}}\right)\left(\frac{m}{10^{-21}\text{eV}}\right)\left(\frac{v}{\text{kms}^{-1}}\right)^{-1}.
\end{equation}
In our fiducial simulation, the peak velocity is $\sim$80km$\textrm{s}^{-1}$, and $\beta=0.11$. At lower velocities, $\beta$ increases, and the assumption $\beta\ll 1$ becomes invalid. However, lower black hole velocities are associated with low-density regions, where  dynamical friction is less significant. Hence, the integrated dynamical friction is dominated by the high-velocity, high-density central regions, where $\beta \ll 1$ is more likely. For larger black hole masses, the assumption $\beta \ll 1$ becomes less appropriate, and poorer correspondence is expected between the semi-analytic model and numerical results.

\begin{figure}[tb]
    \includegraphics[width=\columnwidth]{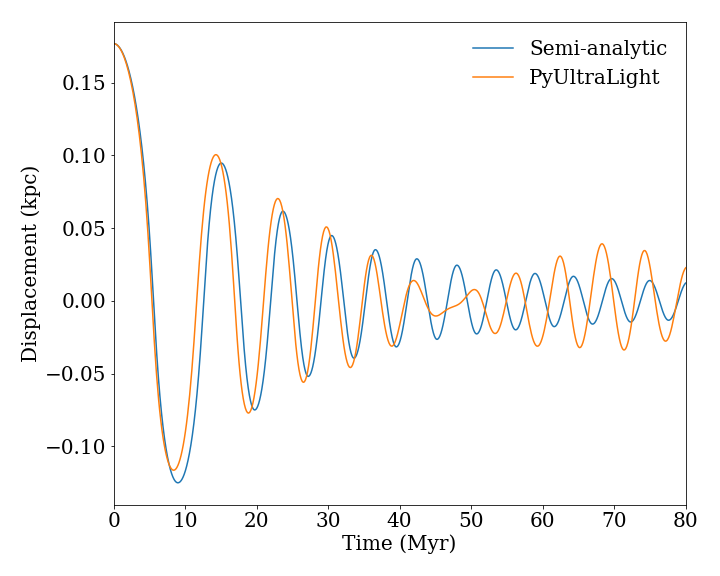}
\\
    \includegraphics[width=\columnwidth]{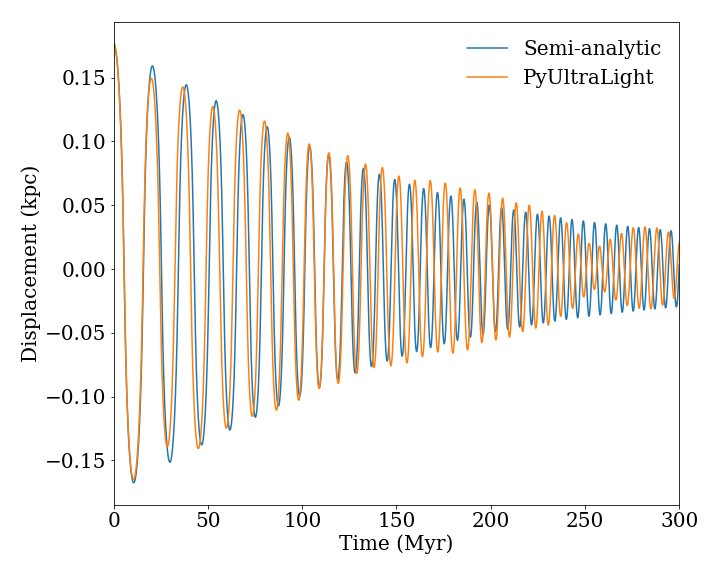}

\caption{Comparison of simulations and the semi-analytic damping model for our fiducial parameters ($m=10^{-21}\text{eV}, M_s = 5.57\times 10^7 \text{M}_\odot$). Top: black hole mass is $\sim$ 7\% of the soliton mass ($\mathcal{M} = 4\times 10^6\text{M}_\odot$). Bottom: black hole mass is 1\% of the soliton mass ($\mathcal{M} = 5.57\times 10^5 \text{M}_\odot$).}
\label{r_vs_sa}
\end{figure}

A key consideration is the specific cutoff, $\tilde{r}$, employed in Equation \ref{cof}. For objects orbiting within virialised halos, $\tilde{r}$ is generally taken to be the radius of the orbit. However, this choice is questionable here, since the radial distance and local density change significantly over time. In particular, by using the radial position of the black hole  as the cutoff we would be causing the (predicted) dynamical friction to vanish in the high-density, high-velocity regime. 

Surprisingly, though, using the radial position as the cutoff yields a reasonable correspondence with the numerical simulations. Figure \ref{r_vs_sa} compares the semi-analytic model with $\tilde{r}$ equal to the black hole radius with our simulations for the fiducial case and a black hole that is $1\%$ of the  soliton mass.

\begin{figure*}
    \centering
    \includegraphics[width=\textwidth]{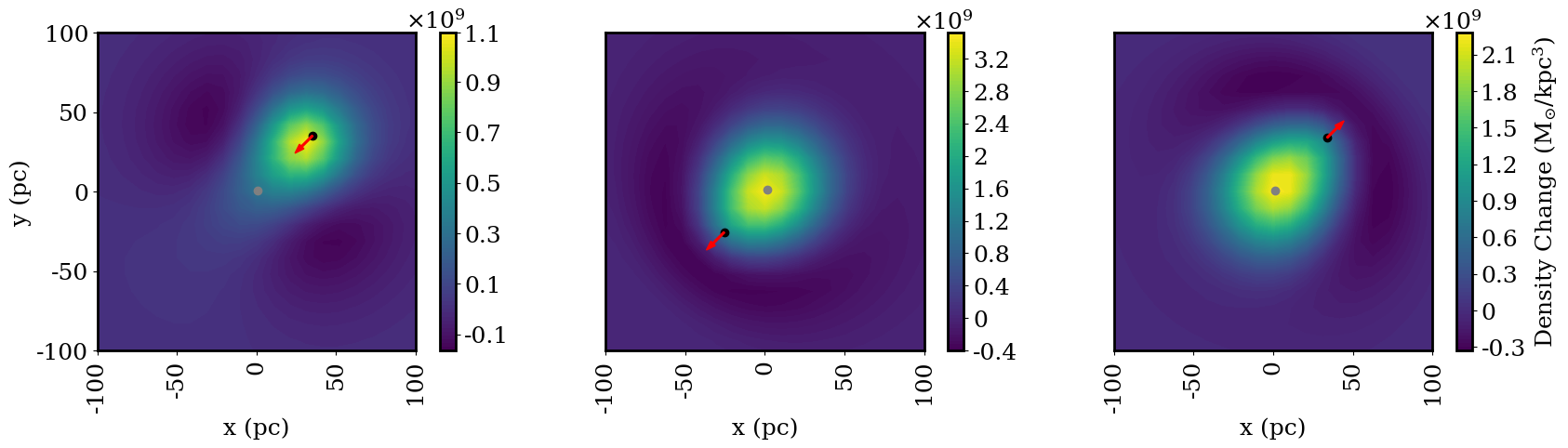}
    \caption{Plots of the net change in ULDM density during a $1\% \text{M}_s$ black hole's passage through the fiducial soliton. The black hole is represented by the black dot, while the ULDM centre of mass is represented by the grey dot. The direction of the black hole's velocity is indicated by the red arrow.  Left: The black hole's initial approach of the soliton centre, showing an overdensity in front of the black hole. Middle: the black hole after its first pass through the centre of mass, demonstrating the appearance of a wake. Right: the black hole after its second pass through the centre of mass.}
    \label{overdensity_0.01}
\end{figure*}

The semi-analytic model matches the full numerical simulations reasonably well, although it slightly underestimates the drag. In fact, the accuracy of match is  surprising, given both  the seemingly unphysical choice of $\tilde{r}$, and because the $\beta \ll 1$ criterion is not rigorously fulfilled.  That said, we found that several potential alternative choices of $\tilde{r}$ actually yield results  less consistent with the simulations. These include setting $\tilde{r}$ equal to the core radius of the soliton, and a position-dependent cutoff such that the soliton density remains approximately constant within the domain of integration. The latter  scheme is reminiscent of the prescription employed in Ref~\cite{2005A&A...431..861J}, which studied drag in a nonuniform matter distribution. 

While the semi-analytic model predicts the integrated effect of dynamical friction relatively well, it is worth investigating whether the assumptions made in its derivation  are indeed representative of the physical processes driving the  damping. Firstly, we can ask whether the distortion of the ULDM density profile by the black hole can be reasonably  described as an overdense wake. Figure \ref{overdensity_0.01} shows the difference between the ULDM density and the initial unperturbed soliton at various times, for the $1\% \text{M}_s$ black hole, centering the unperturbed profile on the ULDM centre of mass.  The left panel shows the relative change in the field configuration when the black hole begins its first pass through the soliton, approaching from the top right with a velocity  indicated by the red arrow. Rather than a distinct overdensity behind the black hole, the soliton appears to have gained a higher-density region in front of the black hole. The field disturbance thus does not resemble the simple trailing wake seen in the scenario of a black hole moving through a uniform background \cite{Wang_2022}.

After the first pass through the soliton, an overdensity resembling a wake can be seen, as shown in the central plot of Figure \ref{overdensity_0.01}, although it is distorted by the intrinsic shape of the soliton itself.  A similar wake arises during subsequent passes, shown in the right hand plot.\footnote{While Figure \ref{overdensity_0.01} corresponds to a $1\%\text{M}_s$ black hole, larger mass black holes cause more significant disruption of the soliton which can be less clearly accommodated by the semi-analytic model.} 

\begin{figure}[tb]
    \centering
    \includegraphics[width=\linewidth]{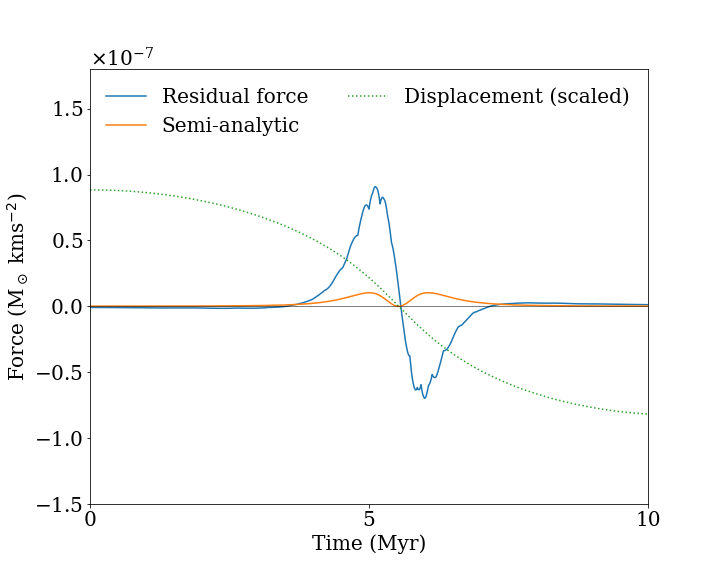}
    \caption{Comparison of the numerical `residual force' with the drag force implied by the semi-analytic model for the $1\%\text{M}_s$ black hole. The black hole's displacement relative to the centre of soliton's centre of mass is also shown for reference, with an arbitrary scaling.}
    \label{residual_vs_sa}
\end{figure}

Overall, the  plots in Figure \ref{overdensity_0.01} suggest that dynamical friction during black hole-soliton interactions is not fully described  by the semi-analytic model.  To test this more quantitatively, we apply the semi-analytic model in reverse. That is, we extract the total force acting on the black hole from our numerical simulations and express this as the  gravitational attraction force due to a spherically symmetric soliton, plus an additional contribution, which we  refer to as the residual force.\footnote{Such a decomposition is of course only technically valid in the case of a spherically symmetric density distribution, an assumption which is increasingly violated at higher black hole mass. Hence, we focus on a $0.01 \text{M}_s$  black hole.} We then compare this to the semi-analytic drag.\footnote{The numerical residual force is computed using the second derivative of a cubic spline interpolation of the black hole displacement. Irregularities in the curve arise from the finite simulation timestep.} Results for the first pass of the $0.01 \text{M}_s$ mass black hole through the soliton centre are shown in Figure \ref{residual_vs_sa}.

\begin{figure*}[tb]
\minipage{0.5\textwidth}
    \includegraphics[scale=0.36, trim={0cm 0cm 1.6cm 0cm}]{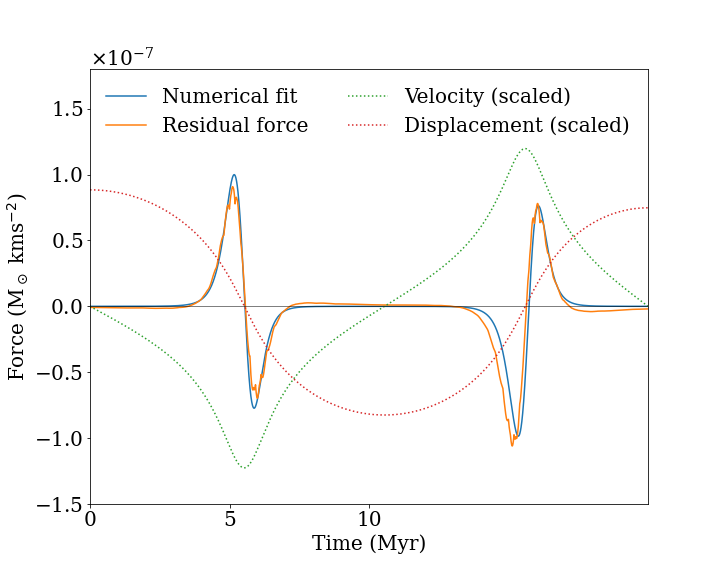}
\endminipage\hfill%
\minipage{0.5\textwidth}
    \includegraphics[scale=0.36, trim={0cm 0cm 1.6cm 0cm}]{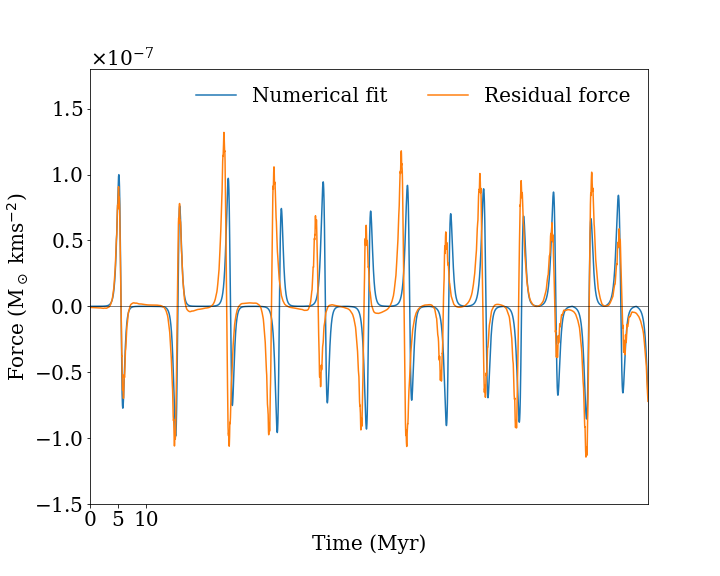}
\endminipage\hfill%
\caption{The empirical fit to the residual force is illustrated for a $0.01\text{M}_s$ black hole.  Left: fit to the residual force for the first two passes through the soliton centre of mass. Right: the same result over multiple passes. Specific parameter values are  $A=1.26\times 10^{-4}\left(\vert v\vert/\text{kms}^{-1}\right)^3$, $B=r/1.6r_c$, $C=13.5$, $D=100$.}
\label{Numerical_fit_compare_1}
\end{figure*}

It is immediately apparent that the  residual force and the semi-analytic drag force are dissimilar, both qualitatively and quantitatively. The magnitude of the residual force is much larger, and while the semi-analytic drag force always acts in the direction opposite to the velocity, the numerical residual force changes sign as the black hole passes through the centre of the soliton, in effect acting as a `negative drag' for part of the trajectory. 

Figure \ref{residual_vs_sa}  demonstrates a significant failure of the semi-analytic model within a regime in which one would expect it to be broadly applicable, namely, in the case where the perturber is only $1\%$ of the mass of the soliton. In more extreme scenarios, it is already known that the semi-analytic model cannot capture the complexities of the perturber-ULDM interactions, but we show here that the semi-analytic model fails for even the simplest of trajectories along the radial direction. 

Surprisingly, however, the integrated effects of the very distinct force curves in Figure \ref{residual_vs_sa} are similar. This coincidence is made possible due to the slight asymmetry between the `positive' and `negative' contributions to the numerical residual force, yielding, on average, a small `positive' contribution over a single passage through the soliton. 

\begin{figure}[tb]
    \includegraphics[width=\columnwidth]{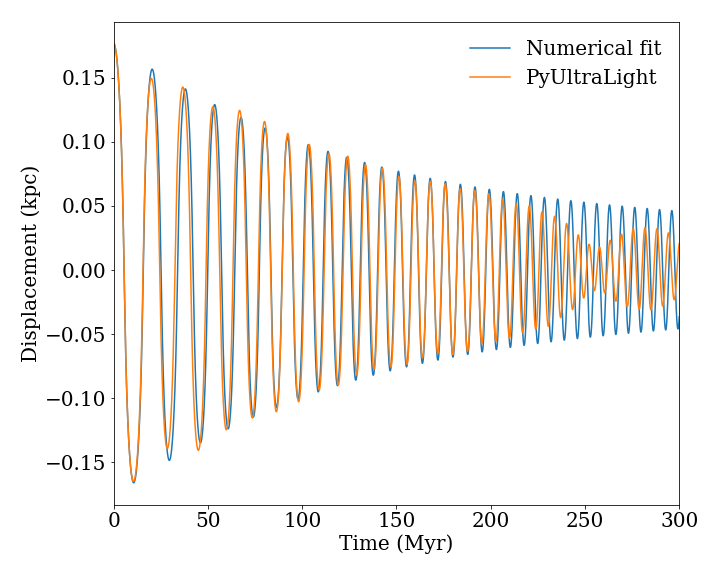}
\caption{Comparison of the numerical evolution of the black hole displacement  with the evolution predicted by the modified semi-analytic model, with the coefficient of friction  given by Equation \ref{gauss_sin}.}
\label{Numerical_fit_compare_2}
\end{figure}

Though the sign change in the numerical residual force may seem counter-intuitive, it has been noted in previous studies that the `drag' force acting on objects travelling through ULDM condensates can  become negative at times, for example when interference crests develop ahead of the test object. In fact, the authors of Ref~\cite{Lancaster:2019mde} argue that such oscillations in the coefficient of friction are a distinctive feature of ULDM, and while they cannot be accommodated in the simple semi-analytic model considered in this work, linear perturbation theory may provide a more accurate estimate.

We find that the force profile takes the same form as shown in Figure \ref{residual_vs_sa} during  subsequent passes through the centre of the soliton. It is therefore interesting to determine whether a numerical fit to the curve shown in Figure \ref{residual_vs_sa} can be used to model the full evolution. In analogy with Equation \ref{Hui_friction} we define
\begin{equation}\label{fit_drag}
    F_{\text{fit}} = \frac{4\pi G^2 {\cal M}^2 \rho}{v^2} \, C_{\text{fit}},
\end{equation}
where $C_{\text{fit}}$ is a dimensionless function of the black hole's position and velocity, $r$ and $v$. The shape of the  curve in Figure \ref{residual_vs_sa} suggests the  form
\begin{equation}\label{gauss_sin}
    C_{\text{fit}} = A\exp\left(-\frac{(B\pm C)^2}{D}\right)\sin\left(B\right),
\end{equation}
where the positive sign in the argument of the exponential applies when $v$ is positive, and vice versa. This purely empirical construction, provides a good fit to the residual force during the first pass. The force derived from this expression for the drag is shown in Figure \ref{Numerical_fit_compare_1}.  The effect of this force on black hole displacement is shown in Figure \ref{Numerical_fit_compare_2}, along with the results from the simulation. We see that using Equation \ref{fit_drag} in the semi-analytic model does indeed yield better agreement with simulation results than Equation \ref{Hui_friction}. The numerical fit to Equation \ref{gauss_sin} is specific to this idealised system, but we anticipate that the dynamical friction in generic ULDM field configurations will also exhibit non-trivial, oscillatory behaviour. 

\section{Conclusion} \label{sec:conclusion}

We have numerically analysed the dynamical friction experienced by a black hole moving on a radial trajectory in a ULDM soliton  and compared our results to analytical predictions. As expected, the loss of the black hole kinetic energy due to dynamical friction becomes more efficient as the mass of the black hole or the ULDM particle increases. Thus, particularly in large galaxies which would be expected to contain larger black holes, dynamical friction from a ULDM core could have a significant effect on black hole dynamics. 

Notably, the efficiency of the dynamical friction is strongly dependent on the mass of the constituent ULDM particle. There are a number of constraints on this  mass, summarised aptly in \cite{Ferreira_2021}. These generally favour higher mass ULDM particles (within their overall range), although black hole superradiance excludes particles with masses above $7\times10^{-20}$eV \cite{Stott_2018}. 

The black hole initially undergoes damped, near-harmonic motion as it makes multiple passes through the soliton. However, the trajectory eventually ``reheats'', and the amplitude of the oscillations increases. This growth appears to be pumped by oscillations in the soliton itself, which result from the transfer of energy from the moving black hole to the ULDM.  This is reminiscent of the ``stone skipping'' trajectories found for black holes in initially circular orbits  \cite{Wang_2022}.  Conversely, treatments that ignore the backreaction on the soliton will understandably predict a monotonic decay  \cite{Annulli_2020}. 

Naively, it might appear that the semi-analytic model of dynamical friction is a reasonably good match for the early stages of the fully non-linear simulations, before the amplitude of the oscillations starts to grow. However,  despite  reasonable  agreement with a simple dynamical friction model, we see that it breaks down when examined in detail. In particular, we find  an apparent `negative friction' as the black hole passes through the centre of the soliton. This arises from the collective response of the soliton to the incoming black hole, in contrast to the simpler overdense wake that acts to damp the motion of a particle moving in a uniform background. We find an empirical fit to the  force profile, but a comprehensive physical description of this phenomenon is not yet available, although a more rigorous perturbative analysis may provide greater insight \cite{Annulli_2020, Lancaster:2019mde}.

This reheating  will be of importance to the final parsec problem \cite{Milosavljevic:2002ht} in ULDM, as this mechanism could prolong (perhaps indefinitely) the time taken for an incoming SMBH to ``sink'' to the centre of a galactic halo. That said, this scenario is highly idealised, since we are considering on a single static soliton with no existing SMBH. Conversely, a realistic situation would  involve the merger of two solitons and two SMBH, and examining this scenario would require a more complex analysis. 

The reheating gives way to an oscillatory radial motion whose amplitude was modulated by a longer term periodicity. This appeared to continue for an arbitrary time, but at some point the results are dominated by numerical effects. While long simulations at high resolution are numerically expensive, and future investigations of this behaviour will benefit from the use of adaptive mesh solvers \cite{Schwabe_2020}.

In summary, we have examined the dynamical friction experienced by a black hole on a radial trajectory within a ULDM soliton. This has shown that the simplest accounts of the  drag do not match the detailed dynamics. Moreover, the long-term motion is oscillatory motion inside a modulated envelope, driven by the excitations of the soliton induced by the passage(s) of the black hole, and this backreaction will be critical to a full understanding of the system. The works helps to elucidate the impact of ULDM solitons on SMBH dynamics and mergers, with relevance to  current and future gravitational wave detectors, including pulsar timing arrays and the LISA space interferometer \cite{LISA:2022kgy}.

\begin{acknowledgments}

We thank Benedikt Eggemeier, Mateja Gosenca, Jens Niemeyer, Nikhil Padmanabhan and Luna Zagorac for useful conversations during the course of this work and acknowledge a useful discussion with Priyamvada Natarajan at the outset of this analysis.   This work is supported by the Marsden Fund of the Royal Society of New Zealand and we acknowledge the use of New Zealand eScience Infrastructure (NeSI) high performance computing
facilities.

\end{acknowledgments}

\bibliography{apssamp}

\end{document}